\documentclass[%
reprint,
superscriptaddress,
 amsmath,amssymb,
 aps,
]{revtex4-2}

\usepackage{hyperref}

\usepackage{graphicx}
\usepackage{dcolumn}
\usepackage{bm}

\begin{document}

\title{Internal Smith-Purcell radiation and its interplay with Cherenkov diffraction radiation in silicon -- a combined time and frequency domain numerical study}

\author{Dmytro Konakhovych}
\author{Damian \'{S}nie\.{z}ek}
\author{Oskar Warmusz}
\affiliation{Institute of Experimental Physics, University of Wroclaw, Plac M.~Borna 9, 50-204 Wroclaw, Poland}
\author{Dylan S.\ Black}
\author{Zhexin Zhao}
\affiliation{Department of Electrical Engineering, Stanford University, 350 Serra Mall, Stanford, California 94305-9505, USA}
\author{R. Joel England}
\affiliation{SLAC National Accelerator Laboratory, 2575 Sand Hill Road, Menlo Park, California 94025, USA}
\author{Andrzej Szczepkowicz}
\affiliation{Institute of Experimental Physics, University of Wroclaw, Plac M.~Borna 9, 50-204 Wroclaw, Poland}
\email{Corresponding author}

\begin{abstract}
We consider radiation generated by an electron travelling parallel to a planar rectangular silicon grating: Smith-Purcell radiation to the vacuum side, internal Smith-Purcell radiation into the dielectric, and Cherenkov radiation into the dielectric. Internal Smith-Purcell radiation dominates over the other two radiation mechanisms in the range where conventional Smith-Purcell radiation is forbidden. This observation may lead to improved design of contactless particle beam monitors.
\end{abstract}

\maketitle

Cherenkov radiation (CR) is a form of light emitted when a charged particle moves through a medium with velocity greater than the speed of light in the medium \cite{1934-Cherenkov, 1937-Frank-Tamm, 1937-Cherenkov}. CR has long been applied in particle physics for particle detection \cite{2004-Akimov}. Recently considerable attention is given to a half-space variant of CR, where a particle moves in vacuum, but parallel to a dielectric medium. This variant of CR \cite{1955-Linhart, 1966-Ulrich, 2000-Takahashi-Shibata} has recently been called Cherenkov diffraction radiation (CDR) and is considered as a new tool for non-invasive (contactless) charged particle beam diagnostics \cite{2018-Kieffer-Bartnik, 2019-Alves-Bergamashi, 2020-Curcio-Bergamashi, 2020-Kieffer-Bartnik, 2020-Lasocha-Lefevre}.

Belonging to the same family of ``Cherenkovian effects'' \cite{1960-diFrancia}, Smith--Purcell radiation (SPR) \cite{1953-Smith-Purcell, 1960-diFrancia, 2010-Potylitsyn-Ryazanov} is the light emitted when a charged particle moves parallel to a periodically corrugated surface. SPR has also been considered for particle beam monitoring \cite{1989-Fernow, 1997-Lampel, 1997-Nguyen, 2001-Doucas-Kimmitt, 2002-Doucas-Kimmitt, 2009-Blackmore-Doucas, 2012-Bartolini-Clarke, 2012-Soong-Byer, 2021-Konoplev-Doucas}. However, the majority of work on SPR was carried out for metal surfaces (like metallic reflective diffraction gratings). In this configuration no radiation enters the solid -- SPR is emitted only to the vacuum side. Although some recent experimental studies of SPR deal with dielectric gratings \cite{2018-Yang-Massuda,2019-Roques-Carmes-Kooi}, they focus on ``conventional'' SPR to the vacuum side.

A recent paper that reports a numerical study of SPR radiation into the vacuum, mentions that for dielectric grating, more energy is transmitted into the grating bulk \cite{2020-Szczepkowicz-Schachter-England}. The authors speculate that this energy could be reflected to the vacuum side using distributed Bragg reflectors. However, this inward radiation, which was considered as loss in the previous literature, could be detected directly with an optical fiber attached to the back of the grating; it could also be coupled to on-chip waveguides as an integrated light source \cite{2019-Roques-Carmes-Kooi}. We call the inward radiation ``internal Smith-Purcell radiation'' (iSPR) (Fig. 1). 

The coexistence of SPR and CDR has been studied for more complicated physical systems: nonlinear materials \cite{2012-Ren-Deng}, and dielectric covered with metallic layer with an array of holes \cite{2012-Zhang-Zhang}; in contrast to these references, here we consider a simple rectangular grating of single optically linear material -- silicon. SPR has also been found to enhance CDR in cylindrical metallic waveguides filled with dielectric with periodically variable radius \cite{2013-Ponomarenko-Ryazanov, 2015-Ponomarenko-Lekomtsev, 2016-Lekomtsev-Aryshev}, which may lead to efficient THz sources; in contrast to this, here we focus on a planar rectangular dielectric grating, which can be nanofabricated for the optical regime using well established silicon wafer processing technology.

In this letter we report a combined time and frequency domain numerical study of iSPR\&CDR in a dielectric planar grating (silicon). We hope that the research initiated here will eventually lead to new, non-invasive beam monitors based on iSPR\&CDR. This perspective appears particularly attractive for the recently developed dielectric laser accelerators \cite{2012-Soong-Byer, 2014-England-Noble, 2019-Yousefi-Schonenberger, 2020-Sapra-Yang}.

Figure 1 shows the planar dielectric grating considered here. We use the Huygens principle \cite{2010-Potylitsyn-Ryazanov} to derive the wavelength--angle relation for iSPR. 
From translational symmetry, the waves emitted at $P$ and $Q$, at the time the electron traverses $P$ and $Q$ respectively, have the same phase.
For constructive interference inside the grating, the time of flight of the electron from $P$ to $Q$ ($\frac{a}{\beta c}$) must differ from radiation time from $P$ to $P'$ ($\frac{a\cos\theta}{c/n}$) by an integer multiple of the field oscillation period ($m \lambda_0 / c$); this leads to the following formula:
\begin{equation}\label{eq-iSPRCDR}
\frac{1}{\beta}-n\cos\theta=\frac{m \lambda_0}{a}
\end{equation}
(see also Refs.~\cite{2012-Ren-Deng, 2015-Ponomarenko-Lekomtsev}).
Alternatively one may note that from the phase matching condition (momentum matching), the momentum mismatch between the electron $\omega/v$ and the plane wave inside the dielectric $n\cos(\theta)\omega/c$ should be compensated by the grating, such that 
$\omega/v - n\cos(\theta)\omega/c = 2\pi m/a$; this condition is equivalent to Eq.~\ref{eq-iSPRCDR}.

In the context of SPR, CDR, and iSPR, one must take care not to confuse the vacuum wavelength $\lambda_0$ with the wavelength in the dielectric $\lambda_0/n$.
For $m\neq0$, Eq.~(\ref{eq-iSPRCDR}) describes iSPR of order $m$ inside a dielectric medium with refractive index $n$; for $n=1$ one obtains the formula for conventional SPR \cite{2010-Potylitsyn-Ryazanov}. For $m=0$,  the constructive interference condition becomes independent of the wavelength and grating period, and Eq.~(\ref{eq-iSPRCDR}) becomes the formula for the Cherenkov angle (see also Ref.~\cite{2015-Ponomarenko-Lekomtsev}). In this way Eq.~(\ref{eq-iSPRCDR}) describes all the three types of radiation: CDR, SPR, iSPR, emphasizing similarities among these ``Cherenkovian effects'' \cite{1960-diFrancia}.

\begin{figure}[htbp]
\label{fig-1}
\centering
\includegraphics[width=0.8\linewidth]{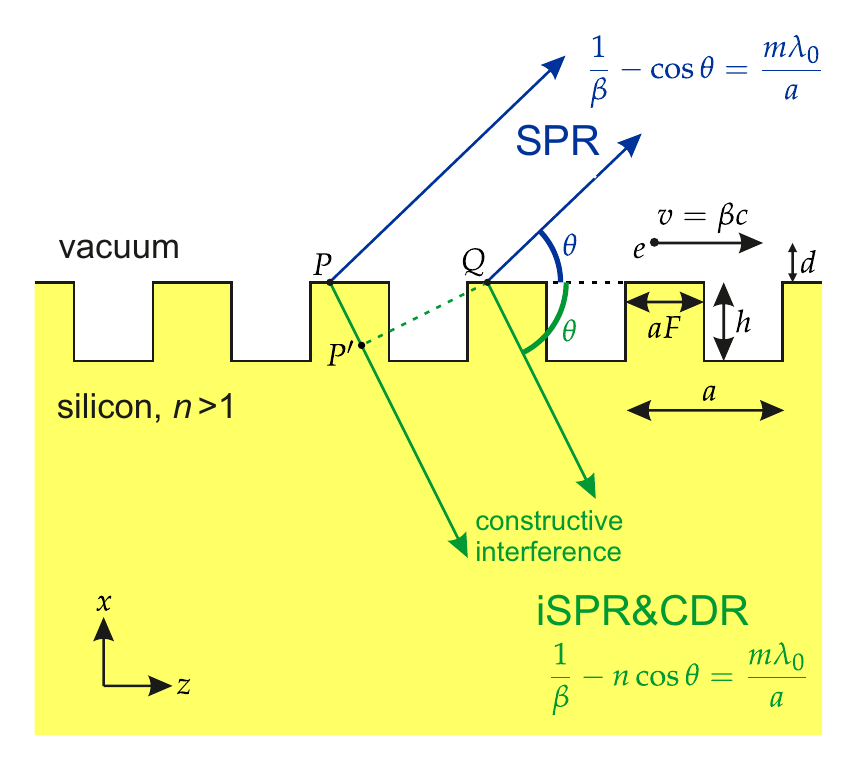}
\caption{Radiation inside a planar silicon grating (iSPR and CDR) and outside (SPR).}
\end{figure}

To study the two kinds of radiation inside silicon, iSPR and CDR, we perform a 3D time-domain simulation (PIC solver in CST Studio Suite), combined with a 2D frequency-domain simulation \cite{2020-Szczepkowicz-Schachter-England} (Comsol Multiphysics). As in Fig.~1, we assume a single electron moving in vacuum parallel to a silicon grating. 
The computations were carried out for the following parameters (refer to Fig.~1): grating period $a=200$~nm, grating duty cycle $F=0.5$ (except Fig.~2b, where $F=1$), grating tooth height $h=175$~nm (however $h$ is varied in Fig.~5), and impact parameter $d=200$~nm (except Fig.~4, where $d = 20$~nm). 
We assume a constant and lossless refractive index (for real silicon the change in refractive index is small in the considered frequency range), $n(\mathrm{Si})=3.91$ (corresponding to first order iSPR in the perpendicular direction at $\lambda_0=609$~nm). 
Electron relative velocity is $\beta=0.328$, corresponding to 30 keV electrons. 
In 3D computations, we assume that the charge is equal to the elementary charge $e$, slightly diffused in the longitudinal and transverse dimensions to avoid PIC solver artifacts (Gaussian distribution with cutoff at radius 200~nm). In 2D computations, we assume a sheet beam Dirac delta pulse with transverse charge density 
$1\ e / \mathrm{1\ nm}$, and analyzed radiation from a corresponding transverse 1 nm strip, as in Refs. \cite{2018-Yang-Massuda, 2020-Szczepkowicz-Schachter-England}.

With the parameters assumed above, Eq.~(\ref{eq-iSPRCDR}) implies that the frequency range for first order iSPR is 
$(215\,\mathrm{THz},\infty)$, 
and for second order iSPR is 
$(431\,\mathrm{THz},\infty)$. 
For comparison, the frequency ranges of conventional SPR are, for first order SPR, $(371\ \mathrm{THz}, 733\ \mathrm{THz})$, 
and for second order SPR 
$(741\ \mathrm{THz}, 1466\ \mathrm{THz})$.

Note that above the threshold velocity for CDR ($n\beta>1$) there is no kinematic upper frequency limit for iSPR, in contrast to SPR to the vacuum side. Also, while the angles for SPR are in the full $(0^{\circ}, 180^{\circ})$ range, the angular range of iSPR is limited by the Cherenkov angle: $(\theta_0,180^{\circ})$, where in our case $\theta_0=38.9^{\circ}$.
Note also that in our case the frequency ranges of radiation orders $m=1$ and $m=2$ are disjoint for SPR and overlapping for iSPR. The general criterion for the $m$-th order to overlap with $(m+1)$-th order is
$\beta > 1/(2m+1)$ for SPR and $n\beta > 1/(2m+1)$ for iSPR.

Figure 2a shows the electromagnetic field resulting from an electron moving above a silicon grating. In addition to the Coulomb field around the electron, iSPR wave oscillation is clearly seen; the dominant wavelength $\lambda_0/n=278$~nm
observed in direction $\hat k_1$ corresponds to vacuum wavelength $\lambda_0=1086$~nm and frequency 276~THz. The measured angle $\theta$ for $\hat k_1$ is $128^{\circ}$; according to Eq.~(\ref{eq-iSPRCDR}), this angle corresponds to first-order iSPR radiation at 275~THz. iSPR radiation at other angles is also  seen, but the field amplitude is smaller.
In the direction marked $\hat k_2$, a CDR shock wave is clearly seen, which, as expected, is a superposition of a wide range of wavelengths. The measured Cherenkov angle $38^{\circ}$ is close to the theoretical value from Eq.~(\ref{eq-iSPRCDR}) obtained for $m=0$: $\theta_0=38.9^{\circ}$. If the grating duty cycle is increased to $F=1$, as shown in Fig.~2b, a plain silicon slab is obtained,
and iSPR disappears as expected.

\begin{figure}[htbp]
\label{fig-2}
\centering
\includegraphics[width=0.95\linewidth]{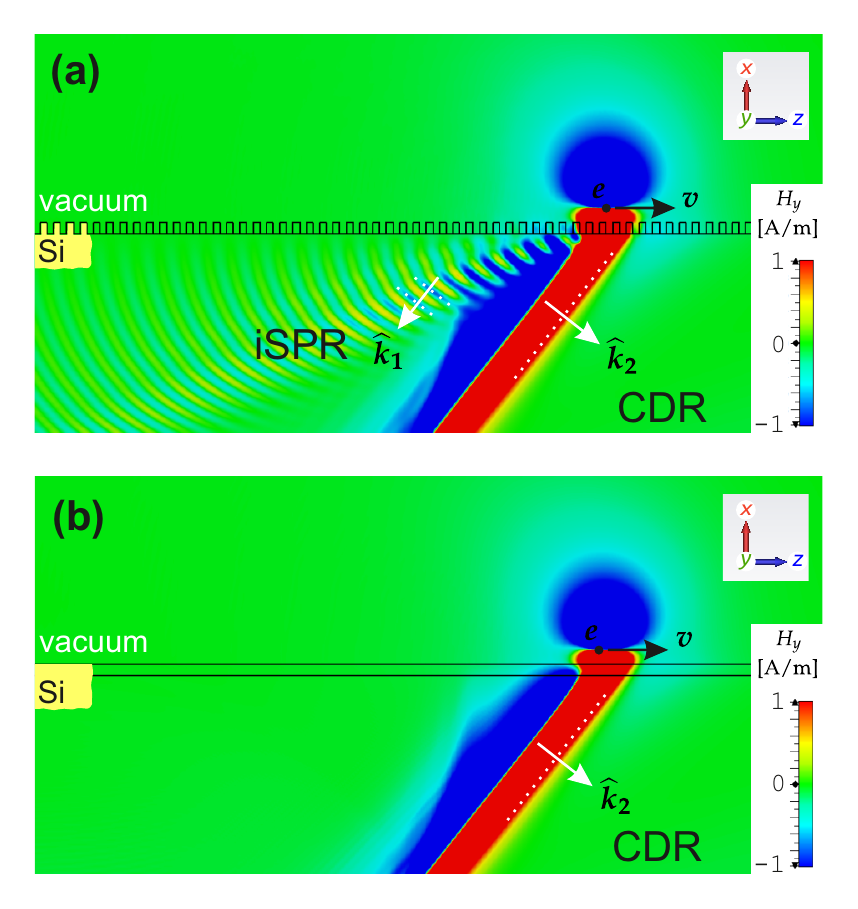}
\caption{(a) Two kinds of radiation inside a silicon grating, iSPR and CDR, in a 3D time-domain simulation. The angles of $\hat k_1$ and $\hat k_2$, relative to $v$, are $128^{\circ}$ and $38^{\circ.}$ Full field is shown, with no frequency filtering. (b) When the grating duty cycle is increased to $F=1$ (planar slab), only CDR remains.}
\end{figure}

Figure 3 shows a corresponding frequency-domain result. The green curve indicates total inward spectral energy density  consisting of iSPR and CDR components, for one grating period. Unfortunately the numerical value of energy cannot provide input for real experiments, because as described above it is a 2D model which assumes a sheet beam pulse (transverse line charge). 
The peak of the iSPR\&CDR curve around 300~THz corresponds to first order iSPR at $\theta_1=120^{\circ}$, which is close to the angle $128^{\circ}$ observed in the time-domain simulation (Fig.~2).
This is first order iSPR radiation; note that at this frequency second order iSPR radiation is kinematically suppressed (Eq.~(\ref{eq-iSPRCDR})). Figure 3 shows also two additional curves for comparison. The blue curve indicates conventional SPR into the vacuum. The threshold for this radiation is 371 THz, so below this limit all radiation energy flows into the grating. It appears that the intensity of iSPR is especially high for the wavelengths for which conventional vacuum SPR is forbidden. The third, red curve, shows CDR radiation for the case of planar silicon slab, corresponding to configuration of Fig. 2b. It is interesting that in the frequency range $(260,340\,\mathrm{THz})$, inward radiation for the grating (iSPR\&CDR) is higher than the corresponding radiation for the planar slab (CDR), despite the fact that in the case of a grating the dielectric is on the average farther away from the electron. 
The iSPR\&CDR from the grating at 308 THz exceeds CDR from a planar slab by a factor of 4. A similar result was obtained in Ref.~\cite{2013-Ponomarenko-Ryazanov} for a different geometry: a cylindrical waveguide filled with dielectric with periodically variable inner radius. This brings hope that beam detectors based on iSPR could be more sensitive than the CDR detectors that are investigated currently \cite{2018-Kieffer-Bartnik, 2019-Alves-Bergamashi, 2020-Curcio-Bergamashi, 2020-Kieffer-Bartnik, 2020-Lasocha-Lefevre}, depending on detection bandwidth.  For example, if particle beam modulation at a particular frequency is to be detected, iSPR signal may considerably exceed CDR signal.

\begin{figure}[htbp]
\label{fig-3}
\centering
\includegraphics[width=0.9\linewidth]{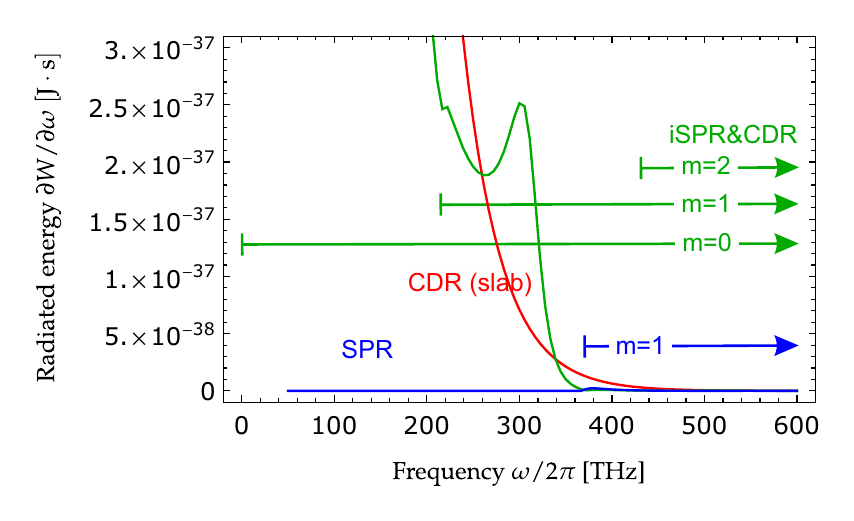}
\caption{A 2D frequency-domain simulation corresponding to the time-domain result in Fig. 2a. Radiated energy per unit frequency from one grating period. Blue line: conventional SPR into the vacuum. Green line: combined iSPR\&CDR in silicon. For comparison, red curve represents pure CDR from a planar silicon slab, corresponding to Fig. 2b. Allowed radiation orders $m$ are indicated, calculated from Eq.~(\ref{eq-iSPRCDR}).}
\end{figure}

Figure 4 illustrates the significance of the impact parameter $d$. When $d$ is reduced from 200~nm (Fig.~3) to 20~nm (Fig.~4), the inward iSPR\&CDR radiation energy (green curve) increases by $\sim$3 orders of magnitude (the same increase is observed for CDR radiation from a planar slab -- red curve). At the same time the iSPR\&CDR peak around 300 THz changes shape and becomes more symmetric. The dependence of the shape of the curve on the impact parameter $d$ could be used for determination of $d$, similarly as in Ref.~\cite{2019-Yevtushenko-Dukhopelnykov} which considers optical beam position monitors based on diffraction radiation from a pair of dielectric nanowires.

\begin{figure}[htbp]
\label{fig-4}
\centering
\includegraphics[width=0.9\linewidth]{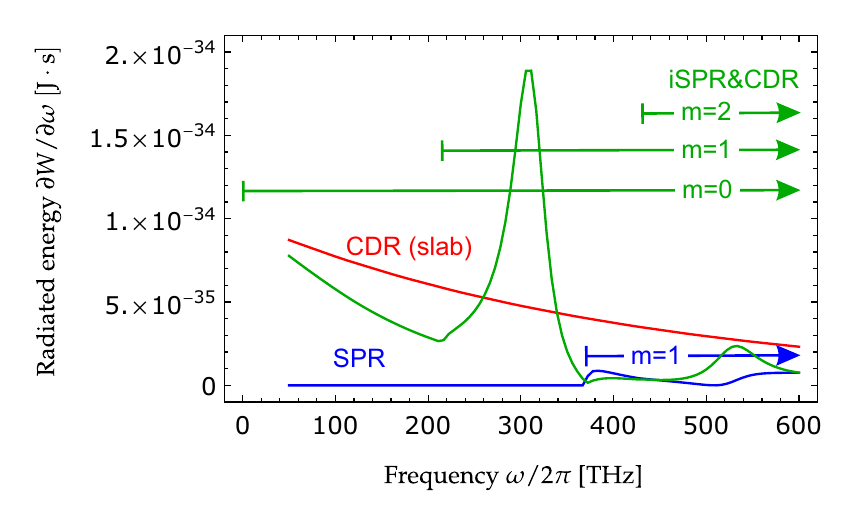}
\caption{Same as Fig. 3, except that the impact parameter is reduced from $d=200$~nm to 20~nm.}
\end{figure}

\begin{figure}[htbp]
\label{fig-5}
\centering
\includegraphics[width=0.9\linewidth]{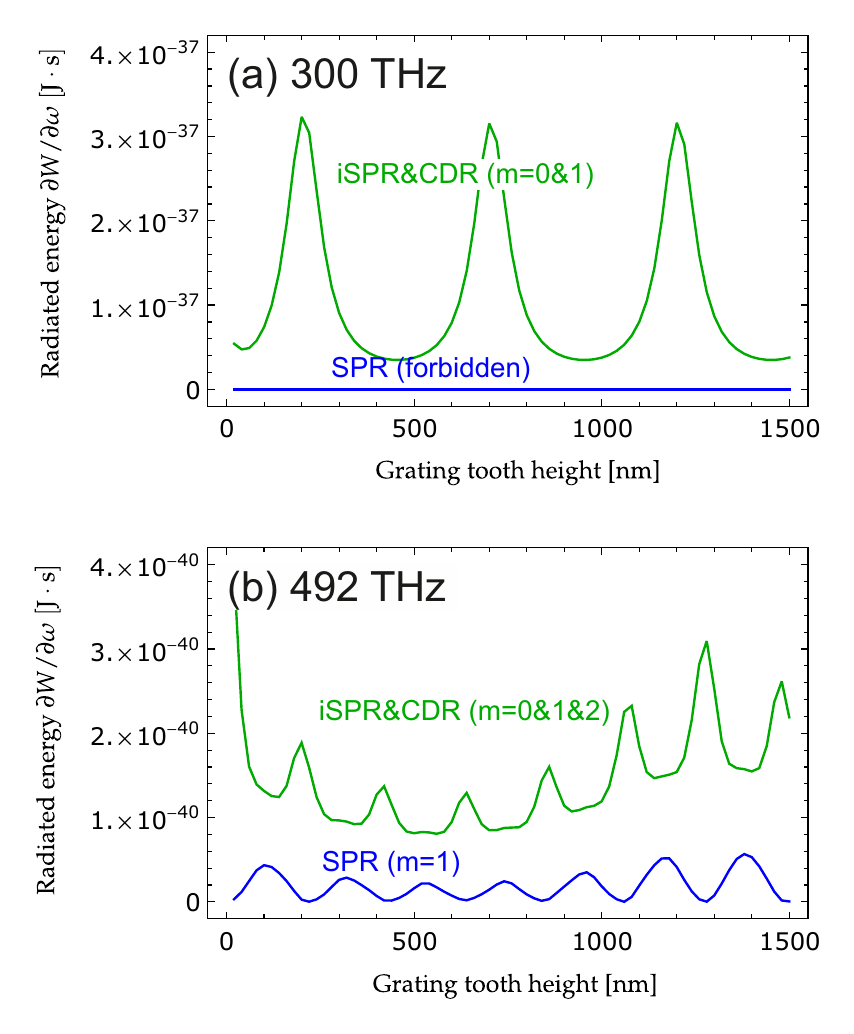}
\caption{A 2D frequency-domain simulation corresponding to the time-domain result in Fig. 2a, except that the grating tooth height $h$ is varied. Vertical axis: radiated energy per unit frequency from one grating period. Blue line: conventional SPR into the vacuum. Green line: combined iSPR\&CDR in silicon. Radiation orders $m$ are indicated.}
\end{figure}

A realistic design of a beam monitor based on iSPR would require optimization of the shape of the dielectric. This optimal shape will depend on the assumed radiation detector type, its geometry and wavelength range. In Fig.~5 we demonstrate how iSPR\&CDR energy at two selected frequencies changes with grating tooth height $h$. At 300 THz only two radiation orders are involved: $m=0$ (CDR) and $m=1$ (iSPR). The emitted energy changes periodically with grating tooth height, similarly as for conventional SPR (see eg. Fig.~6 in Ref. \cite{2020-Szczepkowicz-Schachter-England}), except that it has a constant component connected with CDR. At 492 THz one more radiation order is involved: $m=2$ (iSPR, second order) and consequently the dependence on $h$ is less simple. Note that the maxima for SPR and for iSPR\&CDR occur at different $h$ -- this means that a grating optimized for conventional SPR radiation may not perform well if iSPR is to be detected.

In summary, we investigated numerically the internal Smith-Purcell radiation (iSPR) in a silicon planar grating and its coexistence with Cherenkov diffraction radiation (CDR). We found that that the spectral intensity of iSPR may be several times larger than the spectral intensity of CDR or conventional vacuum Smith-Purcell radiation (SPR). In contrast to SPR, iSPR can be efficiently coupled to an optical fiber or an integrated photonic circuit, without free space propagation of radiation. We found that the intensity of iSPR is especially high for the wavelengths for which SPR is forbidden, and that the optimal geometry for iSPR is different than the optimal geometry for SPR. These findings may lead to new, efficient, non-invasive particle beam monitors for both conventional and laser particle accelerators \cite{2014-England-Noble, 2019-Yousefi-Schonenberger, 2020-Sapra-Yang}
-- perhaps the designs for particle beam monitoring based on conventional SPR 
\cite{1989-Fernow, 1997-Lampel, 1997-Nguyen, 2001-Doucas-Kimmitt, 2002-Doucas-Kimmitt, 2009-Blackmore-Doucas, 2012-Bartolini-Clarke, 2012-Soong-Byer, 2021-Konoplev-Doucas}
or CDR \cite{2018-Kieffer-Bartnik, 2019-Alves-Bergamashi, 2020-Curcio-Bergamashi, 2020-Kieffer-Bartnik, 2020-Lasocha-Lefevre}
could now be adapted for iSPR.

\begin{acknowledgments}
The authors gratefully acknowledge discussions with Yen-Chieh Huang and Levi Sch\"{a}chter.

This research was funded by Gordon and Betty Moore Foundation (GBMF4744).
\end{acknowledgments}

\bibliography{article}


\end{document}